\documentstyle[12pt]{article}
\begin{document}
\pagestyle{empty}
\hspace*{12.4cm}MPT-PhT/2000-21\\
\hspace*{13cm}IU-MSTP/42 \\
\hspace*{13cm}hep-lat/0006008 \\
\hspace*{13cm}June, 2000
\begin{center}
 {\Large\bf Topological Obstruction in Block-spin Transformations}
\end{center}

\vspace*{1cm}
\def\thefootnote{\fnsymbol{footnote}}
\begin{center}
{\sc Takanori Fujiwara${}^1$}\footnote{Permanent address: 
Department of Mathematical Sciences, Ibaraki University,
Mito 310-8512, Japan}, 
{\sc Takuya Hayashi${}^2$,}
{\sc Hiroshi Suzuki${}^{3\ast}$}
and {\sc Ke Wu${}^2$}\footnote{Permanent address: 
Institute of Theoretical Physics, Academia Sinica, P.O.Box 
2735, Beijing 100080, China}
\end{center}
\vspace*{0.2cm}
\begin{center}
{\it ${}^1$Max-Planck-Institut f\"ur Physik, Werner-Heisenberg-Institut, \par
F\"ohringer Ring 6, D-80805 M\"unchen, Germany} \par 
{\it ${}^2$Department of Mathematical Sciences, Ibaraki 
University,
Mito 310-8512, Japan \par
${}^3$ High Energy Group, Abdus Salam ICTP, Trieste, 34014, 
Italy}
\end{center}
\vfill
\begin{center}
{\large\sc Abstract}
\end{center}
\noindent
Block-spin transformations from a fine lattice to a coarse one 
are shown to give rise to a one-to-one correspondence between the 
zero-modes of the Ginsparg-Wilson Dirac operators. The index is then 
preserved under the blocking process. Such a one-to-one correspondence 
is violated and the block-spin transformation becomes necessarily 
ill-defined when the absolute value of the index is larger than $2rN$, 
where $N$ is the number of the sites on the coarse lattice and $r$ is 
the dimension of the gauge group representation of the fermion variables. 



\newpage
\setcounter{page}{1}
\pagestyle{plain}
The technique of block-spin transformation is familiar in 
quantum field theory and statistical physics. It can be used to reduce 
the number of the degrees of freedom of a system without losing the 
characteristic features of the original system such as the behaviors 
under the symmetry transformations. When it is applied to a fermion system 
described by an action that is chirally symmetric and bilinear in 
the fermion fields, the Dirac operator of the transformed system 
turns out to satisfy the so-called Ginsparg-Wilson (GW) relation 
as the remnant of the chiral symmetry \cite{GW}. The recent discoveries 
of lattice Dirac operators 
satisfying the GW relation \cite{has}, which are free from the 
species doubling while keeping all the desired properties expected 
for lattice Dirac operators such as the classical continuum limit and 
the locality \cite{HJL}, have triggered a new development in lattice 
chiral gauge theories \cite{Reviews}. The relationship among the exact 
chiral symmetry on the lattice \cite{Lus}, the chiral anomaly and the index 
theorem \cite{HLN} has been clarified by using the algebraic 
properties of the GW relation. The perturbative evaluation of the 
chiral anomaly in the overlap formalism has been carried out in 
ref. \cite{KY}. See also refs. \cite{Adams} for further evaluations 
of the chiral anomaly in the continuum limit. The index theorem 
of the GW Dirac operator and the chiral anomaly of the overlap Dirac 
operator are also investigated in refs. \cite{Fujik,chiu,Fujik2}. 

Since any gauge field configuration on the lattice can be continuously 
deformed into the topologically trivial one, the space of the link variables 
is topologically trivial. But it can be topologically disconnected 
by excising the exceptional configurations \cite{lus4}. The restriction 
of the link variables to a topologically disconnected space may also emerge 
on physical grounds. For instance the overlap Dirac operator is 
considered to 
be smooth and local on some restricted gauge field configuration space 
with a nontrivial topological structure \cite{HJL}, otherwise one 
would not obtain any nontrivial topological invariants. The index theorem 
\cite{HLN,Lus} relates the index of the GW Dirac operator to the integral 
of the lattice chiral anomaly, a function of the lattice gauge field, 
into which the topological information of the gauge field configuration 
space is transcribed. In the case of abelian 
gauge theory on an infinite regular lattice it is possible to find the 
explicit expression for the lattice chiral anomaly in terms of the lattice 
gauge field by noting its locality, gauge invariance and topological 
invariance \cite{Lus2,FSW}. Furthermore, it can be interpreted as the 
Chern character of the lattice abelian gauge theory \cite{FSW,FSW2}. 
In the case of nonabelian theories, however, no such explicit expression 
for the lattice chiral anomaly is available at present and the 
relationship 
between the topology of the gauge field configuration and the index of 
the GW Dirac operator is still not so clear for strictly finite lattice 
spacing \cite{Adams2}.

In this note we investigate the indices of the GW Dirac operators 
within the framework of the block-spin approach \cite{GW} and try to 
shed some light on the connection of the indices with the topology 
of the gauge field. We show that there is a one-to-one correspondence 
between the zero-modes of the GW Dirac operators related to each other 
by a block-spin transformation and, hence, the index is preserved 
under the transformation. Since the index is also invariant under 
arbitrary continuous variations of the gauge field, some of the 
zero-modes must be topologically stable 
if the index is nonvanishing. The number of topologically stable
zero-modes is just the absolute value of the index. 
Such a correspondence between the zero-modes and the persistence of the 
index are already argued in ref. \cite{Yamada}, where the author 
considered the block-spin transformation of a chirally symmetric 
continuum Dirac theory by assuming that the resulting lattice is fine 
enough for the shape of the zero-modes of the original continuum 
theory to be well preserved. We will generalize his argument from a 
more general setting and analyze what happens in the block-spin approach 
when the index of the original GW Dirac operator is very large. We shall see 
that the block-spin transformation to a coarse lattice cannot 
be well-defined due to the excess of the topologically stable zero-modes 
when the index becomes extremely large. 

Let us consider a fermion system $\{\phi,\bar \phi\}$ on a finite 
euclidean lattice $\Lambda_0$ with a Dirac operator ${\cal D}$ satisfying 
the GW relation
\begin{eqnarray}
  \label{eq:GWrelfl}
  \gamma_5{\cal D}+{\cal D}\gamma_5
  =2{\cal D}\gamma_5{\cal R}{\cal D}~,
\end{eqnarray}
where the matrix ${\cal R}$ is assumed to be regular, hermitian and 
possessing positive 
definite eigenvalues so that $\sqrt{{\cal R}}$ is unambiguously defined. 
We also assume that ${\cal R}$ is local on $\Lambda_0$ and commutes with 
any Dirac matrix. We suppose that the fermion system is coupled 
to an external gauge field and the GW Dirac operator 
${\cal D}$ satisfies $\gamma_5{\cal D}^\dagger\gamma_5={\cal D}$, 
hence $\gamma_5{\cal D}$ is hermitian. 

If the gauge field configuration carries a nonvanishing topological 
charge, there appear zero-modes of the Dirac operator ${\cal D}$. 
By the GW relation they can be chosen to have definite chiralities. 
Let $n_+$ and $n_-$ be, respectively, the number of the positive and 
negative chirality zero-modes, then the index is given 
by \cite{HLN,Lus,chiu,Fujik2}
\begin{eqnarray}
  \label{eq:inddef}
  n_+-n_-={\rm Tr}\gamma_5(1-{\cal R}{\cal D})~, 
\end{eqnarray}
where ${\rm Tr}$ implies the summation over the lattice coordinates 
as well as the trace of the spin and internal indices. 
This can be seen by considering the eigenvalue problem $\gamma_5
\hat{\cal D}\phi_\lambda=\lambda\phi_\lambda$ of 
the hermitian operator $\gamma_5\hat{\cal D}\equiv\gamma_5
\sqrt{{\cal R}}{\cal D}
\sqrt{{\cal R}}$  \cite{Fujik2}, 
where $\phi_\lambda$ is assumed to be normalized and the eigenvalue 
$\lambda$ satisfies $-1\le\lambda\le1$ 
as can be seen from $||\gamma_5(1-\hat{\cal D})\phi_\lambda||^2
=1-\lambda^2\ge0$. Since $\hat{\cal D}$ satisfies 
$\gamma_5\hat{\cal D}\gamma_5(1-\hat{\cal D})
=-\gamma_5(1-\hat{\cal D})\gamma_5\hat{\cal D}$, we see that 
$\phi_{-\lambda}\propto\gamma_5(1-\hat{\cal D})\phi_\lambda$ is 
linearly independent of $\phi_\lambda$ for $\lambda\ne0,\pm1$. 
This also implies the orthogonality 
$(\phi_\lambda,\gamma_5(1-\hat{\cal D})\phi_\lambda)=0$. 
The eigenmodes for $\lambda=\pm1$ are necessarily chiral with chirality 
$\pm$ since they satisfy $\gamma_5(1-\hat{\cal D})
\phi_{\pm1}=(\gamma_5\mp1)\phi_{\pm1}=0$. Hence in the computation 
of the index 
\begin{eqnarray}
  \label{eq:comptr}
  {\rm Tr}\gamma_5(1-{\cal R}{\cal D})={\rm Tr}\gamma_5(1-\hat{\cal D})
  =\sum_\lambda(\phi_\lambda,\gamma_5(1-\hat{\cal D})\phi_\lambda)~,
\end{eqnarray}
only the zero-modes $\phi_0$ contribute to the index, giving the relation 
(\ref{eq:inddef}).

Since the index (\ref{eq:inddef}) is a topological invariant as can 
be directly verified by using the GW relation (\ref{eq:GWrelfl}), 
the $|n_+-n_-|$ zero-modes are stable under arbitrary local 
variations of the gauge potential. This fact is well-known in the continuum 
theory and holds true also in the lattice theory. We now show 
that the index is also invariant under block-spin 
transformations of the fermion variables. 

We suppose that $\Lambda$ is a coarse sublattice of $\Lambda_0$ and the 
blocked variables are given by 
\begin{eqnarray}
  \label{eq:bv}
  (\rho\phi)_X\equiv\sum_{x\in\Lambda_0}\rho_{Xx}\phi_x~,
  \qquad (\bar\phi\rho^\dagger)_X\equiv
  \sum_{x\in\Lambda_0}\bar\phi_x\rho^\dagger_{xX}~, 
  \qquad (X\in\Lambda)~.
\end{eqnarray}
We assume that $\rho$ is local and commutes with any Dirac matrix. 
For a reason that will be clear later we also assume $\rho\rho^\dagger$ 
to be regular on $\Lambda$. To maintain the gauge invariance we choose 
the averaging functions $\rho$ and $\rho^\dagger$ to be gauge covariant 
though this is not necessary in showing the one-to-one correspondence 
between the zero-modes. In what follows we shall extensively use 
index free matrix notation. 

Let us consider a fermion system 
$\{\psi,\bar\psi\}$ on a lattice $\Lambda$, and then define the block-spin 
transformation of the fermion variables by the following path integral
\begin{eqnarray}
  \label{eq:bst}
  {\rm e}^{-\bar\psi D\psi}
  =\int d\phi d\bar\phi
  {\rm e}^{-(\bar\psi-\bar\phi\rho^\dagger)
    \alpha(\psi-\rho\phi)-\bar\phi{\cal D}\phi}~,
\end{eqnarray}
where the matrix $\alpha$ is assumed to be hermitian, local and regular on 
$\Lambda$ and proportional to the unit matrix in the Dirac space. We 
also assume that $\alpha^{-1}$ is local. The simplest choice consistent 
with the gauge symmetry is for $\alpha$ to be proportional to the unit matrix 
on $\Lambda$. We leave the external gauge field unchanged, and so the Dirac 
operator $D$ on the coarse lattice $\Lambda$ still depends on the gauge 
field on the fine lattice $\Lambda_0$. 

It is straightforward to show that the Dirac operator $D$ on the 
coarse lattice is given by 
\begin{eqnarray}
  \label{eq:dopcl}
  D=\alpha-\alpha\rho {\cal M}^{-1}\rho^\dagger\alpha~,
\end{eqnarray}
where ${\cal M}$ is defined by the Dirac operator ${\cal D}$ on the fine 
lattice $\Lambda_0$ as
\begin{eqnarray}
  \label{eq:M}
  {\cal M}={\cal D}+\rho^\dagger\alpha\rho~.
\end{eqnarray}
Furthermore by using (\ref{eq:GWrelfl}) and (\ref{eq:dopcl}) the Dirac 
operator $D$ on the coarse lattice can be shown to satisfy the following 
GW relation
\begin{eqnarray}
  \label{eq:GWrelcl}
  D\gamma_5+\gamma_5D=2D\gamma_5 R D~,
\end{eqnarray}
where $R$ is given by
\begin{eqnarray}
  \label{eq:R}
  R=\alpha^{-1}+\rho{\cal R}\rho^\dagger~.
\end{eqnarray}
These results are already noted in ref. \cite{HLN}.

That the index (\ref{eq:inddef}) is conserved under the block-spin 
transformation can be seen as follows: 
the fermion action on the lattice $\Lambda_0$ is invariant 
under the infinitesimal chiral transformation \cite{Lus}
\begin{eqnarray}
  \label{eq:ict}
  \phi'=\Bigl\{1+
  i\epsilon\gamma_5(1-{\cal R}{\cal D})\Bigr\}\phi~, 
  \qquad
  \bar\phi'=\bar\phi\Bigl\{1+
  i\epsilon(1-{\cal D}{\cal R})\gamma_5\Bigr\}~. 
\end{eqnarray}
The fermion measure, however, gives rise to the Jacobian 
\cite{Fuj}
\begin{eqnarray}
  \label{eq:jacb}
  d\phi'd\bar\phi'=d\phi d\bar\phi
  \Bigl\{1-2i\epsilon{\rm Tr}\gamma_5(1-{\cal R}{\cal D})\Bigr\}~.
\end{eqnarray}
The term proportional to $\epsilon$ in the Jacobian must be canceled 
by a term arising from the blocking kernel 
$(\bar\psi-\bar\phi\rho^\dagger)\alpha(\psi-\rho\phi)$ . The significance 
of the 
Jacobian is also noted in refs. \cite{SU,Yamada}. By applying the chiral 
transformation (\ref{eq:ict}) to the rhs of (\ref{eq:bst}) and 
retaining only the first order terms in $\epsilon$ it is straightforward 
to show that the fermion system on the coarse lattice $\Lambda$ must 
satisfy
\begin{eqnarray}
  \label{eq:WT}
  \Biggl\{{\rm Tr}
  \gamma_5\frac{\partial}{\partial\bar\psi}\Biggl[
  \Biggl(\bar\psi-\frac{\partial}{\partial\psi}R\Biggr)\Biggr]
  +\frac{\partial}{\partial\psi}
  \gamma_5\Biggl(\psi+R\frac{\partial}{\partial\bar\psi}\Biggr)
  -2{\rm Tr}\gamma_5(1-{\cal R}{\cal D})\Biggr\}
  {\rm e}^{-\bar\psi D\psi}=0~,
\end{eqnarray}
where the derivatives act on everything on their right. 
This gives the GW relation (\ref{eq:GWrelcl}) and the conservation 
of the index
\begin{eqnarray}
  \label{eq:ii}
    {\rm Tr}\gamma_5(1-RD)&=&{\rm Tr}\gamma_5(1-{\cal R}{\cal D})~.
\end{eqnarray}
This result can also be obtained directly from (\ref{eq:dopcl}) and 
(\ref{eq:M}) together with the GW relation (\ref{eq:GWrelfl}) \cite{HLN}. 
We note the following identity 
\begin{eqnarray}
  \label{eq:1-rd}
  1-RD=\rho(1-{\cal R}{\cal D}){\cal M}^{-1}\rho^\dagger\alpha~.
\end{eqnarray}
Taking the trace of this expression and using 
${\cal M}^{-1}\rho^\dagger\alpha\rho=1-{\cal M}^{-1}{\cal D}$, we get 
\begin{eqnarray}
  \label{eq:trg5}
  {\rm Tr}\gamma_5(1-RD)={\rm Tr}\gamma_5(1-{\cal R}{\cal D})
  -{\rm Tr}{\cal D}\gamma_5(1-{\cal R}{\cal D}){\cal M}^{-1}~.
\end{eqnarray}
The second term on the rhs of this expression, however, can be shown 
to vanish by using ${\rm Tr}{\cal D}\gamma_5{\cal M}^{-1}
={\rm Tr}\gamma_5{\cal D}
{\cal M}^{-1}$ and the GW relation (\ref{eq:GWrelfl}). We thus see that 
(\ref{eq:ii}) is reproduced. 

The index relation (\ref{eq:ii}) should not be confused with the 
topological invariance of the index. The latter is concerned with the 
continuous change of the gauge field, whereas the former implies the 
persistence of the index under the discrete transformations.\footnote{%
The block-spin transformations are discrete since the degrees of freedom 
cannot be continuously changed.} This suggests that there is
some relationship between the zero-modes of the two GW Dirac operators. 
We can indeed show a much stronger 
statement than (\ref{eq:ii}) that there is a chirality preserving 
one-to-one correspondence between the zero-modes; the persistence
of the index is an immediate consequence of this fact. 

To show the one-to-one correspondence we first note the following identities
\begin{eqnarray}
  \label{eq:opids}
  D\rho=\alpha\rho {\cal M}^{-1}{\cal D}~, \qquad
  \rho^\dagger D={\cal D}{\cal M}^{-1}\rho^\dagger\alpha~.
\end{eqnarray}
The first relation implies that $\varphi\equiv\rho\Phi$ is a zero-mode 
of $D$ if $\Phi$ is a nonvanishing zero-mode of ${\cal D}$ \cite{Yamada}. 
That $\varphi$
is nonvanishing can be shown by noting the following identity 
\begin{eqnarray}
  \label{eq:phisnv}
  \Phi=(1-{\cal M}^{-1}{\cal D})\Phi={\cal M}^{-1}\rho^\dagger\alpha\varphi~.
\end{eqnarray}
On the other hand we see from the second relation of (\ref{eq:opids}) that 
$\Phi={\cal M}^{-1}\rho^\dagger\alpha\varphi$ is a zero-mode of 
${\cal D}$ if $\varphi$ is a nonvanishing zero-mode of $D$. Since 
$\rho\rho^\dagger$ and $\alpha$ are nonsingular by assumption, 
this relation can be 
inverted as $\varphi=\rho\Phi$ and $\Phi$ cannot be nonvanishing 
if $\varphi\ne0$. We thus obtain the one-to-one correspondence between 
the zero-modes 
\begin{eqnarray}
  \label{eq:1-1}
  \varphi=\rho\Phi\quad\longleftrightarrow\quad\Phi={\cal M}^{-1}\rho^\dagger
  \alpha\varphi~.
\end{eqnarray}
Note that the chirality is preserved under the mapping between the 
zero-modes. From $\Phi$ to $\varphi$ this is obvious 
since $\rho$ commutes with $\gamma_5$. Conversely from $\varphi$ to 
$\Phi$ one can 
confirm this by noting the relation 
$\gamma_5\Phi={\cal M}^{-1}\rho^\dagger\alpha
\gamma_5\varphi$ for $D\varphi=0$. We thus find that under the block-spin 
transformation a zero-mode of ${\cal D}$ on the fine lattice with definite 
chirality is mapped to a zero-mode of $D$ on the coarse lattice with the 
same chirality and vice versa. Therefore $n_+$ and $n_-$ are preserved 
separately under the block-spin transformation. 

We now take gauge field configurations with very large topological 
charge on the fine lattice $\Lambda_0$ and consider a block-spin 
transformation to the very coarse lattice $\Lambda$. The question is how 
coarse may $\Lambda$ be? As we have shown, the index is preserved if 
the block-spin transformation exists. However, there are at most $2rN$ 
zero-modes of $D$ on $\Lambda$, where $N$ stands for the number of 
the sites on $\Lambda$ and $r$ is the dimension of the 
gauge group representation of the fermion variables. The factor $2$ 
comes from the four dimensional representation of the Dirac matrices. 
The reduction from $4$ to $2$ can be understood as follows; we 
note that there must appear $N_\pm$ eigenmodes $\omega_\pm$ defined by 
$\gamma_5RD\omega_\pm=\pm\omega_\pm$ satisfying the chirality sum rule 
$n_++N_+=n_-+N_-$ due to ${\rm Tr}\gamma_5=0$ \cite{chiu} and 
the obvious inequality 
$n_++n_-+N_++N_-\le 4rN$ coming from the size of the GW Dirac operator 
$D$. These imply that $n_\pm$ cannot be larger than $4rN/2=2rN$. 
On the other hand, since $\max\{n_+,n_-\}$ 
cannot be smaller than $|n_+-n_-|$, we find that the index satisfies 
$|n_+-n_-|\le\max\{n_+,n_-\}\le2rN$ whenever the block-spin 
transformation from $\Lambda_0$ to $\Lambda$ is well-defined. 
In the extreme case of $n_+=N_-=2rN$ and $n_-=N_+=0$ the GW Dirac 
operator $D$ is simply given by $R^{-1}(1-\gamma_5)/2$. We see that 
in this extreme case all the propagating degrees disappear on the coarse 
lattice. 

Conversely, this implies that it is impossible to 
find such a block-spin transformation if $\Lambda$ is so coarse and 
\begin{eqnarray}
  \label{eq:bind}
  |{\rm Tr}\gamma_5(1-{\cal R}{\cal D})|>2rN
\end{eqnarray}
is satisfied. In other words 
the block-spin transformation becomes ill-defined if the number 
of the sites on the coarse lattice multiplied by $2r$ is less than the 
absolute value of the index of the GW Dirac operator on the fine lattice. 
We now show that this indeed occurs. 

Let $N_0$ be the number of the sites on the fine lattice $\Lambda_0$. 
Then $\rho$ can be represented as a matrix with $4rN$ rows and $4rN_0$ 
columns including spin and internal degrees of freedom. Since 
$\rho\rho^\dagger$ and $\alpha$ are assumed to be 
regular $4rN\times 4rN$ matrices, a zero-mode of $\rho$ is always a 
zero-mode of $\rho^\dagger\alpha\rho$ and vice versa. In general 
$\rho$ possesses $4r(N_0-N)$ zero-modes since the maximal rank of such 
matrices is $4rN$. We denote the zero-modes of $\rho$ with definite 
chirality by $u_\pm^{(i)}$ ($i=1,\cdots,2r(N_0-N)$) and the nonzero-modes 
by $v_\pm^{(p)}$ ($p=1,\cdots,2rN$). They are assumed to be linearly 
independent and satisfy $\gamma_5u_\pm^{(i)}=\pm u_\pm^{(i)}$ and  
$\gamma_5v_\pm^{(p)}=\pm v_\pm^{(p)}$. 

On the other hand let us suppose that the Dirac operator ${\cal D}$ 
on the fine lattice $\Lambda_0$ possesses $n_++n_-$ zero-modes 
$\Phi_\pm^{(a)}$ ($a=1,\cdots,n_\pm$) with $\gamma_5\Phi_\pm^{(a)}
=\pm\Phi_\pm^{(a)}$ and $4rN_0-n_+-n_-$ nonzero-modes. Since 
$u_\pm^{(i)}$ and $v_\pm^{(p)}$ form a basis 
for the fermion variables on $\Lambda_0$, $\Phi_\pm^{(a)}$ can be 
expressed as
\begin{eqnarray}
  \label{eq:lc}
  \Phi_\pm^{(a)}=\sum_{i=1}^{2r(N_0-N)}c^{(\pm)}_{ai}u_\pm^{(i)}
  +\sum_{p=1}^{2rN}d^{(\pm)}_{ap}v_\pm^{(p)}~,
\end{eqnarray}
where $c^{(\pm)}_{ai}$ and $d^{(\pm)}_{ap}$ are some constants. The matrix 
$d^{(\pm)}\equiv(d^{(\pm)}_{ap})$ has $n_\pm$ rows and $2rN$ columns and 
the maximal rank of $d^{(\pm)}$ is $\max\{n_\pm,2rN\}$. If $n_+$ is larger 
than $2rN$ and $d^{(+)}$ is of the maximal rank $2rN$, then there 
exist $n_+-2rN$ sets of $n_+$ constants $C_a^{(s)}$ 
($s=1,\cdots,n_+-2rN$, $a=1,\cdots,n_+$) satisfying 
$\displaystyle{\sum_{a=1}^{n_+}C_a^{(s)}d_{ap}^{(+)}=0}$ for $p=1,\cdots,2rN$. 
The existence of such constants in turn implies that 
certain combinations of the zero-modes of ${\cal D}$ are simultaneously
zero-modes of $\rho$. They are explicitly given by
\begin{eqnarray}
  \label{eq:sez}
  \Psi_+^{(s)}=\sum_{a=1}^{n_+} C^{(s)}_a\Phi_+^{(a)}
  =\sum_{i=1}^{2r(N_0-N)}\sum_{a=1}^{n_+} C^{(s)}_ac^{(+)}_{ai}u_+^{(i)}
\end{eqnarray}
and satisfy ${\cal M}\Psi_+^{(s)}=0$. We thus find that ${\cal M}$ 
must have at least 
$n_+-2rN$ zero-modes if $n_+>2rN$ and, hence, ${\cal M}^{-1}$ does not exist. 
A similar thing also happens in the case $n_->2rN$. In our argument showing 
the persistence of the index it is tacitly assumed that ${\cal M}$ is 
regular and that ${\cal M}^{-1}$ exists. If $\Lambda$ becomes very coarse 
and $\max\{n_+,n_-\}>2rN$ is satisfied, then ${\cal M}$ starts to have 
zero-modes and the regularity of ${\cal M}$ is lost. This implies 
that the block-spin transformation (\ref{eq:bst}) becomes ill-defined 
for gauge field configurations with $\max\{n_+,n_-\}>2rN$. 
This is precisely what happens in the block-spin transformation for a 
given gauge field configuration with a very large topological winding. 
There may occur, however, accidental 
zero-modes of ${\cal D}$ that are not stable under arbitrary local 
variations of the gauge potential and the numbers of the zero-modes 
$n_\pm$ in general may jump for some gauge field configurations within 
the same connected component. Nevertheless we can give a topologically 
invariant statement by noting that the number of topologically stable 
zero-modes of ${\cal D}$ is just the absolute value of the index: 
it is impossible to carry out the block-spin transformation if 
(\ref{eq:bind}) is satisfied. 

In the remainder of this note we consider the case that the 
background gauge field has a smooth continuum limit as $\Lambda_0$ 
approaches to the continuum and give the explicit expression of the 
index of the GW Dirac operator $D$ on $\Lambda$ in terms of the 
smooth background gauge field. This corresponds to considering 
the block-spin transformation of the continuum theory.\footnote{%
In the continuum theory the path integral (\ref{eq:bst}) is assumed 
to be regularized, say, by using the gauge covariant mode cut-off 
\cite{Yamada}.}

The GW Dirac operator ${\cal D}$ is supposed to approach the Dirac 
operator $iD\hskip -.25cm/\equiv i\gamma_\mu(\partial_\mu+A_\mu)$ 
of the continuum theory, where $A_\mu$ is the smooth background 
gauge potential. By invoking the Atiyah-Singer index theorem \cite{AS} 
we see that the index is given explicitly in terms of $A_\mu$ as 
\begin{eqnarray}
  \label{eq:indrel}
  {\rm Tr}\gamma_5(1-RD)=\frac{1}{32\pi^2}\int d^4x
  \epsilon_{\mu\nu\rho\sigma}{\rm tr}F_{\mu\nu}F_{\rho\sigma}~.
\end{eqnarray}
where $F_{\mu\nu}=\partial_\mu A_\nu-\partial_\nu A_\mu+[A_\mu,A_\nu]$ 
is the field strength. It is possible to show this in the present 
context by evaluating the index of $D$ given by (\ref{eq:dopcl}) 
and (\ref{eq:M}) with ${\cal D}_{xy}=\langle x|iD\hskip -.25cm/|y\rangle
=iD\hskip -.25cm/(x)\delta_{xy}$ and ${\cal R}=0$. The index can be 
expressed as  
\begin{eqnarray}
  \label{eq:cind}
  {\rm Tr}\gamma_5(1-RD)={\rm Tr}\gamma_5\rho{\cal M}^{-1}
  \rho^\dagger\alpha
  ={\rm Tr}\gamma_5{\cal V}
  ({\cal D}+{\cal V})^{-1}~,
\end{eqnarray}
where use has been made of (\ref{eq:1-rd}) and we 
have introduced ${\cal V}\equiv\rho^\dagger\alpha\rho$. But 
the rhs of this expression is independent of ${\cal V}$ 
(as far as ${\cal V}$ commutes with $\gamma_5$) as can be 
verified from the identity\footnote{%
This can be derived from $\gamma_5{\cal D}+{\cal D}\gamma_5=0$ 
and $\gamma_5{\cal V}-{\cal V}\gamma_5=0$.} 
\begin{eqnarray}
  \label{eq:idtity}
  \{\gamma_5(1-2{\cal V}({\cal D}+{\cal V})^{-1})\}^2=1~. 
\end{eqnarray}
The operator ${\cal V}$ can be 
continuously changed to any operator unless 
$({\cal D}+{\cal V})^{-1}$ 
becomes singular. In particular we may consider a deformation of 
${\cal V}$ from $\rho^\dagger\alpha\rho$ to $M_0\delta_{xy}$, where 
$M_0$ is a sufficiently large constant. We thus obtain 
\begin{eqnarray}
  \label{eq:limindD}
  {\rm Tr}\:\gamma_5(1-RD)
  =\lim_{M_0\rightarrow\infty}
  {\rm Tr}\gamma_5\frac{M_0}{iD\hskip -.27cm /+M_0}
  =\int d^4x \lim_{M_0\rightarrow\infty}
  \langle x|{\rm tr}\Biggl(\gamma_5\frac{M_0}{iD\hskip -.27cm /+M_0}
  \Biggr)|x\rangle~,
\end{eqnarray}
where ${\rm tr}$ stands for the ordinary trace over the spin and internal 
indices and we have taken the infinite $M_0$ limit for simplicity.  
The integrand of the last expression is nothing but the chiral anomaly in 
the case of Pauli-Villars regularization. We thus obtain (\ref{eq:indrel}). 

That the index of $D$ coincides with that of the continuum theory is not 
so surprising since $D$ still depends on the smooth gauge field $A_\mu$ 
of the continuum theory through ${\cal D}$. In this sense the index 
(\ref{eq:indrel}) should be considered to be related to the topology 
of the continuum gauge fields. Nevertheless, it is possible to apply 
(\ref{eq:indrel}) for the evaluation of the index of the GW Dirac 
operator with the gauge field being defined not in the continuum but on 
the lattice if one can find a smoothly interpolated gauge potential 
in the continuum from the lattice gauge field. The GW Dirac operator 
$D$ is still given by (\ref{eq:dopcl}) and (\ref{eq:M}) but it now 
depends only on the gauge field on the lattice. In this case the 
index of the lattice theory coincides with that of the continuum 
theory. This is quite nontrivial and is not to be confused with the classical 
continuum limit discussed above. Any lattice approximation 
may truncate some topological information of the continuum gauge 
field and such coincidence of the indices between the lattice 
theory and the corresponding continuum limit is not guaranteed 
in general.

We may summarize as follows. The block-spin transformation 
(\ref{eq:bst}) makes sense in general and induces a chirality 
preserving one-to-one mapping between the zero-modes of the GW 
Dirac operators if the number of zero-modes of ${\cal D}$ 
satisfies $n_\pm\le 2rN$. For a suitable choice of $\alpha$ and $\rho$, 
${\cal M}$ has no zero-mode and the inverse ${\cal M}^{-1}$ is 
well-defined. However, if $\max\{n_+,n_-\}>2rN$, some of the 
zero-modes of ${\cal D}$ become also zero-modes of ${\cal M}$ 
for any choice of $\alpha$ and $\rho$. The functional integral 
(\ref{eq:bst}) is vanishing there and it becomes impossible to 
define the Dirac operator $D$ for the fermion system on the coarse 
lattice $\Lambda$. When the absolute value of the index is larger 
than $2rN$, an excess of zero-modes occurs for any gauge 
field configuration within the same topologically connected 
component. Though (\ref{eq:bind}) is rarely satisfied in practical 
situations, it is very interesting to note that the block-spin 
transformations from a fine lattice with a very large index to 
coarser and coarser lattices cannot be carried out successively 
arbitrarily. They become inevitably ill-defined at some stage where 
(\ref{eq:bind}) is satisfied. 

\vskip .3cm
We would like to thank Peter Weisz for valuable comments and 
careful reading of the manuscript. 
This work was completed during the time when two of us 
(T.F. and H.S.) attended 
the Ringberg workshop. They would like to thank Martin L\"uscher, 
Erhard Seiler and Peter Weisz for the kind hospitality. 
T.F. thank Ryuji Uchino for discussions at the early stage of this 
work. He would also like to express his thanks to Dieter Maison for the 
kind hospitality during his stay at Max Planck Institute. K.W. 
is very grateful to Fumihiko Sakata for the warm hospitality and 
to Faculty of Science of Ibaraki University for the financial 
support during his visit at Ibaraki University.


\newpage
\begin{thebibliography}{99}
\bibitem{GW} P.H. Ginsparg and K.G. Wilson, Phys. Rev. {\bf D25} 
(1982) 2649.
\bibitem{has} P. Hasenfratz, Nucl. Phys. (Proc. Suppl.) {\bf 63} 
(1998) 53; 
Nucl. Phys. {\bf B525} (1998) 401;
\newline
H. Neuberger, Phys. Lett. {\bf B417} (1998) 141; {\bf B427} (1998) 
353.
\bibitem{HJL} P. Hern\'andez, K. Jansen and M. L\"uscher, 
Nucl. Phys. {\bf B552} (1999) 363;
\newline
H. Neuberger, Phys. Rev. {\bf D61} (2000) 085015.
\bibitem{Reviews} M. L\"uscher, Nucl. Phys. {\bf B549} (1999) 
295; {\bf B568} (2000) 162; 
\newline
For reviews, see F. Niedermayer, Nucl. Phys. (Proc. 
Suppl.) {\bf 73} (1999) 105;
\newline
M. L\"uscher, Talk given at the 
International Symposium on Lattice Field Theory, Pisa, 
hep-lat/9909150.
\bibitem{Lus} M. L\"uscher, Phys. Lett. {\bf B428} (1998) 342.
\bibitem{HLN} P. Hasenfratz, V. Laliena and F. Niedermayer, 
Phys. Lett. {\bf B427} (1998) 125.
\bibitem{KY} Y. Kikukawa and A. Yamada, Phys. Lett. {\bf B448} (1999) 265.
\bibitem{Adams} D.H. Adams, hep-lat/9812003;
\newline
H. Suzuki, Prog. Theor. Phys. {\bf 102} (1999) 141. 
\bibitem{Fujik} K. Fujikawa, Nucl. Phys. {\bf B546} (1999) 480.
\bibitem{chiu} T.W. Chiu, Phys. Rev. {\bf D58} (1998) 074511.
\bibitem{Fujik2} K. Fujikawa, Phys. Rev. {\bf D60} (1999) 074505.
\bibitem{lus4} M. L\"uscher, Commun. Math. Phys. {\bf 85} (1982) 39;
\newline
A. Phillips, Ann. Phys. {\bf 161} (1985) 399;
\newline
A.S. Kronfeld, M.L. Laursen, G. Schierholz and 
U.-J. Wiese, Nucl. Phys. 
{\bf B292} (1987) 330;
\newline
M. G\"ockeler, A.S. Kronfeld, M.L. Laursen, G. Schierholz and 
U.-J. Wiese, Nucl. Phys. {\bf B292} (1987) 349;
\newline
M. L\"uscher, Nucl. Phys. {\bf B549} (1999) 295.
\bibitem{Lus2} M. L\"uscher, Nucl. Phys. {\bf B538} (1999) 515.
\bibitem{FSW} T. Fujiwara, H. Suzuki and K. Wu, Phys. Lett. 
{\bf B463} (1999) 63; Nucl. Phys. {\bf B569} (2000) 643; 
IU-MSTP/37, hep-lat/9910030. 
\bibitem{FSW2} T. Fujiwara, H. Suzuki and K. Wu, 
IU-MSTP/39, hep-lat/0001029.
\bibitem{Adams2} D.H. Adams, hep-lat/0001014; hep-lat/0004015.
\bibitem{Yamada} A. Yamada, hep-lat/9911017.
\bibitem{Fuj} K. Fujikawa, Phys. Rev. Lett. {\bf 42} (1979) 1195;
Phys. Rev. {\bf D21} (1980) 2848; 
{\bf D22} (1980) 1499 (E).
\bibitem{SU} H. So and N. Ukita, Phys. Lett. {\bf B457} (1999) 314.
\bibitem{AS} M. Atiyah and I. Singer, Ann. Math. {\bf 87} (1968) 484.
\end{thebibliography}
\end{document}